\documentclass[twocolumn,
showpacs,preprintnumbers,amsmath,amssymb,floatfix,aps]{revtex4}

\usepackage[dvips]{graphicx}% Include figure files
\usepackage{dcolumn}% Align table columns on decimal point
\usepackage{bm}% bold math
\begin{document}

%\preprint{APS/123-QED}

%\email{antezza@science.unitn.it}

%\date{\today}

%\maketitle

%\documentstyle[preprint,aps]{revtex}
%%%%%%%%%%%%%%%%%%%%%%%%%%%%%%%%%%%%%%%%%%%%%%%%%%%%%%%%%%%%%%%%%%%%%%%%%%%%%%%%%%%%%%%%%%%%%%%%%%%%%%%%%%%%%%%%%%%%%%%%%%%%

 %\draft

%\begin{document}
\title{Why and when the Minkowski's stress tensor can be used in the problem of
Casimir force acting on bodies embedded in media}
\author{L.P. Pitaevskii}
\date{28.05.2005}
%\maketitle
\email{lev@science.unitn.it}
\affiliation{
Dipartimento di Fisica, Universit\`a di Trento
and Istituto Nazionale
per la Fisica della Materia INFM-BEC , I-38050 Povo, Trento, Italy \\
Kapitza Institute for Physical Problems, ul. Kosygina 2, 119334 Moskow,
Russia}
\begin{abstract}
It is shown that the criticism \cite{Welsch} of the
Dzyaloshinskii-Lifshitz-Pitaevskii theory of the van der Waals-Casimir
forces inside a medium is based on misunderstandings. It is explained why and
at which conditions one can use the  ''Minkowski-like '' stress tensor for
calculations of the forces. The reason, why approach of \cite{Welsch} is
incorrect, is discussed.
\end{abstract}
\maketitle
This notice is a comment on a recently published paper \cite{Welsch}.
The authors of this paper criticize the existing theory \cite{DP,DLP,DLP2} of
the van der Waals-Casimir forces inside a dielectric fluid. In particular
they express doubt about using of  the ''Minkowski's stress tensor'' which
in their opinion has been ''taken for granted without justifications''. The
aim of this letter is to show that this criticism is based on
misunderstandings.

The theory of Refs. \cite{DP,DLP,DLP2} is, indeed, an extension of the Lifshitz
theory of the van der Waals - Casimir forces in vacuum. As the authors Ref. 
\cite{Welsch} correctly notice ''Lifshitz himself did not address nonempty
interspaces in his seminal article \cite{Lif}''. The reason was that the
Lifshitz approach is based on averaging of the stress tensor in vacuum with
respect to electromagnetic fluctuations. This method cannot be used in a
medium, because there is no general expression for the stress tensor of a
non-stationary electromagnetic field inside an absorptive medium. The
problem, however, can be solved for the equilibrium radiation. The tensor of
van der Waals-Casimir forces was for the first time obtained in \cite{DP} in
terms of the Matsubara Green's functions of electromagnetic field by
summation of a proper set of Feynman diagrams for the free energy and its
variation with respect to the density. The general expression for the stress
tensor for a fluid with $\mu =1$  ( in CGSE units) is 
\begin{gather}
\sigma_{ik}=-P_{0}\delta _{ik}-\frac{\hbar T}{2\pi }\Big \{
 \sum_{n=-\infty}^{\infty}\Big(
\varepsilon D_{ik}^{E}+D_{ik}^{H}\nonumber \\
-\frac{1}{2}D_{ii}^{E}\Big [ \varepsilon
-\rho \Big ( \frac{\partial \varepsilon }{\partial \rho }\Big ) _{T}\Big ]
\delta _{ik}-\frac{1}{2}D_{ii}^{H}\delta _{ik}\Big ) \Big \} ,  \label{DP}
\end{gather}
where $D_{ik}^{E}$ and $D_{ik}^{H}$  are the Green's functions associated
with the
components of electric and magnetic fields, $\varepsilon =\varepsilon \left(
\rho ,T,i\zeta _{n}\right) ,\zeta _{n}=2nT/\hbar $ and $P_{0}\left( \rho
,T\right) $ is the pressure as a function of density and temperature in
absence of electric field. Equation $\left( \ref{DP}\right) $ assumes the
system to be in thermal, but still not in {\em mechanical }equilibrium. Notice
that the tensor $\left( \ref{DP}\right) $ is not an analog of the
Minkowski's tensor, but an analog of the famous Abraham stress tensor of the
stationary electromagnetic field in a dielectric fluid (see, for example, 
\cite{LL8} \S \S\ 15, 35, 75): 
\begin{gather}
\sigma_{ik}^{A}=-P_{0}\delta _{ik}+
\frac{\varepsilon E_{i}E_{k}+H_{i}H_{k}}{4\pi }%
\nonumber \\
-\frac{E^{2}}{8\pi }\left[ \varepsilon -\rho \left( \frac{\partial
\varepsilon }{\partial \rho }\right) _{T}\right] \delta _{ik}-\frac{H^{2}}{%
8\pi }\delta _{ik}.  \label{Abr}
\end{gather}
This equation was derived by M. Abraham about 1909 and, in my opinion, is
one of the most important results of the electrodynamics of continuous media.

The general equation $\left( \ref{DP}\right) $ can be simplified in concrete
experimental situations. An important simplification of the tensor $\left( 
\ref{DP}\right) $ (and analogously of $\left( \ref{Abr}\right) $) can be
achieved by {\em assuming} that the fluid, as it occurs in typical
experimental situations, is also in {\em mechanical equilibrium}.  The
force acting per unit of fluid volume 
is $f_{i}=\partial \sigma_{ik}/\partial
x_{k}$. With the help of the Maxwell
equations this expression can be transformed  into the form
\begin{equation}
f_{i}=-\frac{\partial P_{0}}{\partial x_{i}}-\rho \frac{\partial }{\partial
x_{i}}\left[ \frac{\hbar T}{4\pi }\sum\limits_{n}D_{ii}^{E}\left( \frac{%
\partial \varepsilon }{\partial \rho }\right) _{T}\right] .  \label{force}
\end{equation}
The condition  of {\em  mechanical equilibrium } means that the force 
$f_{i}=0$. Let the fluid has {\em uniform density } in the
absence of the field. Using $\rho =const$ and $f_i=0$ in $\left( \ref{force}%
\right) $, one obtains the condition of the equilibrium: 
\begin{equation}
P_{0}+\frac{\hbar T}{4\pi }\sum\limits_{n}D_{ii}^{E}\rho \left( \frac{%
\partial \varepsilon }{\partial \rho }\right) _{T}=const.  \label{mech}
\end{equation}
This equation implies that a part of the stress tensor $\left( \ref{DP}%
\right) $ is constant through the fluid, being a uniform compressing or
expanding pressure. This part can be omitted in many problems, for example
for the calculation of the full force, acting on a body, embedded in the fluid.
Subtracting the constant tensor $\left[ -P_{0}-\frac{\hbar T}{4\pi }%
\sum\limits_{n}D_{ii}^{E}\rho \left( \frac{\partial \varepsilon }{\partial
\rho }\right) _{T}\right] \delta _{ik}$ from $\left( \ref{DP}\right) $, one
obtains the ''Minkowski-like'' tensor 
\begin{equation}
\sigma_{ik}^{M}=-\frac{\hbar T}{2\pi }\left\{ \sum_{n}\left( \varepsilon
D_{ik}^{E}+D_{ik}^{H}-\frac{1}{2}\varepsilon D_{ii}^{E}\delta _{ik}-\frac{1}{%
2}D_{ii}^{E}\delta _{ik}\right) \right\}   \label{VWM}
\end{equation}
which was correctly used in \cite{DLP}  for calculation of the force
between bodies, separated by a dielectric fluid {\em in mechanical
equilibrium}. Hence, tensor $\left( \ref{VWM}\right) $  has not been
''taken for granted without justifications'' as the authors 
of Ref. \cite{Welsch}
say, but has been {\em derived} in \cite{DLP} under well-defined
conditions, namely the conditions of mechanical equilibrium.

Notice that $\partial \sigma_{ik}^{M}/\partial x_{k}=0$ and hence $\oint
\sigma_{ik}^{M}dS_{k}=0$ for integration over any closed surface, surrounding a
volume of a uniform fluid, just due to the fact that in mechanical
equilibrium electromagnetic forces are compensated by gradient of pressure.
This compensation results in a small change of density of the liquid, which
can be determined using the equation of equilibrium $\left( \ref{mech}%
\right) $. In this connection it is strange that the authors of 
Ref. \cite{Welsch}
consider as ''very paradoxical'' the result ''that the force acting on any
slice of material within interspace vanishes identically''. If  it would
not vanish, the force would accelerate the slice, in obvious contradiction
with the assumption of mechanical equilibrium. 

I hope that these remarks clearly show that criticism \cite{Welsch} about 
the use
of the Minkowski-like tensor for the calculation of van der Waals-Casimir
forces is based on a misunderstanding.

In conclusion I would also like to comment about the method developed in 
Ref. \cite{Welsch}. 
The authors of this paper suggested to use, instead of the theory \cite
{DP,DLP,DLP2}, a new one based on the following 
stress tensor, averaged with respect
to fluctuations (see equation (12) and (54) of \cite{Welsch}): 
\begin{equation}
T_{ik}=\frac{\left\langle E_{i}E_{k}+B_{i}B_{k}\right\rangle }{4\pi }-%
\frac{\left\langle E^{2}+B^{2}\right\rangle }{8\pi }\delta _{ik}.  \label{RW}
\end{equation}
I have rewritten (54) of \cite{Welsch} in CGSE units and in tensor
notation to simplify the comparison with the old results. Authors believe that
''use of $T$, which is formally the same as the stress tensor in microscopic
electrodynamics, is {\em always} correct''. I think that the tensor $\left( 
\ref{RW}\right) $ is {\em incorrect}
 and not only because this equation does not
coincide with the correct equation $\left( \ref{DP}\right) $.

The key point is that equation $\left( \ref{RW}\right) $ takes into account only
a part of the microscopical stress tensor. In their derivation the authors,
in a completely arbitrary way, omitted  contribution to the tensor
 from the ''kinetic'' term 
$-\sum_{a}m_{a}v_{ai}v_{ak}$ , where the sum is taken over all
particles in unit of volume of the medium  and $v_{ai}$ 
are the particles velocities. 
(See, for example, \cite{LL2}, \S 33, here 
I consider the non-relativistic case.) 
However, an electromagnetic field changes the motion of the particles.
Thus the kinetic contribution also depends on the field, and corresponding
field-depending terms must be included, as well the pressure term,
in the stress tensor. One can trace
this contribution in calculations of the stress tensor in plasma, where the
problem can be explicitly solved and where 
the kinetic term plays a crucial role (see \cite{PP}).

I thank M. Antezza who attracted my attention to this problem and 
F. Dalfovo for critical reading of the text.


\begin{thebibliography}{99}
%\begin{references}
\bibitem{Welsch}  C. Raabe and D.-G. Welsch, Phys. Rev. A {\bf 71}, 013814
(2005).

\bibitem{DP}  I.E. Dzyaloshinskii and L.P. Pitaevskii, Sov. Phys. JETP {\bf 9%
}, 1282 (1959).{\bf \ }

\bibitem{DLP}  I.E. Dzyaloshinskii, E.M. Lifshitz and L.P. Pitaevskii, Sov.
Phys. JETP {\bf 10}, 161 (1960).

\bibitem{DLP2}  I.E. Dzyaloshinskii, E.M. Lifshitz and L.P. Pitaevskii, Adv.
Phys. {\bf 10}, 165 (1961).

\bibitem{Lif}  E.M. Lifshitz, Sov.
Phys. JETP {\bf 2}, 73 (1956).

\bibitem{LL8}  L.D. Landau and E.M. Lifshitz, {\em Electrodynamics of
Continuous Media}, Pergamon Press, Oxford, 1984.

\bibitem{LL2}  L.D. Landau and E.M. Lifshitz, {\em The Classical
Theory of Fields}, Pergamon Press, Oxford, 1975.

\bibitem{PP}  V.I. Perel and Ya.M. Pinskii, Sov. Phys. JETP {\bf 27}, 1014
(1968).
%\end{references}
\end{thebibliography}
\end{document}